\documentclass[a4paper,11pt]{article}

\pdfoutput=1

\usepackage{a4wide}
\usepackage{graphicx}
\usepackage[usenames]{color}
\usepackage{amssymb}

\usepackage{amsmath}
\usepackage{verbatim}
\usepackage{units}

\usepackage{hyperref}

\begin{document}

\begin{center}   
\textbf{\LARGE From lab to landscape-scale experiments for the morphodynamics of sand dunes}

\vspace*{0.2cm}

P. \textsc{Claudin}$^a$, S. \textsc{Courrech du Pont}$^b$ and C. \textsc{Narteau}$^c$
\end{center}

{\small
\noindent
$^a$ {Physique et M\'ecanique des Milieux H\'et\'erog\`enes, CNRS -- ESPCI -- PSL Research Univ. -- Sorbonne Univ. -- Univ. Paris Cit\'e, Paris, France.}\\
$^b$ {Laboratoire Mati\`ere et Syst\`emes Complexes, Univ. Paris Cit\'e, CNRS, Paris, France.}\\
$^c$ {Institut de Physique du Globe de Paris, Univ. Paris Cit\'e, CNRS, Paris, France.}
}

\begin{abstract}
We review the main processes that drive the morphodynamics of dunes, i.e. their growth in height, migration and elongation, and emphasise the contribution of experiments to the understanding of these mechanisms. The main control parameters are the sediment flux $Q$ and the saturation length $L_{\rm sat}$ associated with the spatial relaxation of the flux towards the transport capacity. The other relevant quantities are essentially dimensionless: fluid response to a bed perturbation, dune geometry (orientation, aspect ratio), transport ratios under multi-directional wind regimes. We argue that laboratory experiments dealing with sedimentary bedforms in water flows are good analogues to study the morphodynamics of aeolian dunes at reduced length and time scales, as $L_{\rm sat}$ and $L_{\rm sat}^2/Q$ are expected to be smaller for subaqueous bedload. Besides, dune shape and dynamics are mainly governed by flow and boundary conditions, independent of the transport mode. We discuss different experimental set-ups and results, especially concerning dune pattern orientation and dune interaction. Under natural wind regimes in terrestrial deserts, we show the potential of field experiments in which the control of initial and boundary conditions allows for the quantification of all the relevant mechanisms involved in dune growth. We emphasise the general agreement between observations, measurements and theoretical predictions, which indicates a robust comprehension of the underlying processes. This understanding can serve as a foundation for further investigations, including the interpretation of dune landscapes and the resolution of inverse problems.
\end{abstract}

\vspace*{0.5cm}
\begin{center}
Comptes-Rendus Physique \textbf{25}, pp 1--29 (2024).\\
\href{https://doi.org/10.5802/crphys.203}{\texttt{doi.org/10.5802/crphys.203}}\\
Contribution to the Special Issue\\
`Geophysical and astrophysical fluid dynamics in the laboratory'.
\end{center}
\vspace*{0.5cm}

\section{Introduction}
\label{intro}
Aeolian sand dunes display magnificent sedimentary patterns in various environments \cite{Lore14,Dini17,Gunn22}: terrestrial deserts \cite{Bree79} and coastal areas \cite{Hesp13a}, but also planetary surfaces such as those of Mars \cite{Cutt73,Bour10,Lapo18}, Titan \cite{Lore06,Rade08,Rodr13}, Venus \cite{Gree84, Basi85, Weit94}, Pluto \cite{Telf18} and comet 67P \cite{Jia17}. The shape, size and orientation of these dunes are determined by various factors, including the prevailing wind regime and the availability of sediment, as well as the hydrodynamic regimes \cite{Cour24}. On Earth, individual dunes span length scales from 10 to $10^3\;{\rm m}$, and they can interact with each other in dune fields of hundreds of kilometres \cite{Bree79}. Small dunes emerge or react quickly to consistent winds over a few days, while large dunes and dune fields integrate wind regimes over long times. This integrative behaviour allows for a rich variety of shapes and makes it difficult to interpret a dune landscape in all its complexity. Dunes are thus witnesses of past winds and they have been used to constrain global climate models on the planetary bodies where they are observed. Performing comparative planetology in dune geomorphology is also a challenge due to the significant variability in environmental parameters.

\begin{figure}[t!]
\begin{center}
\includegraphics[width=0.9\linewidth]{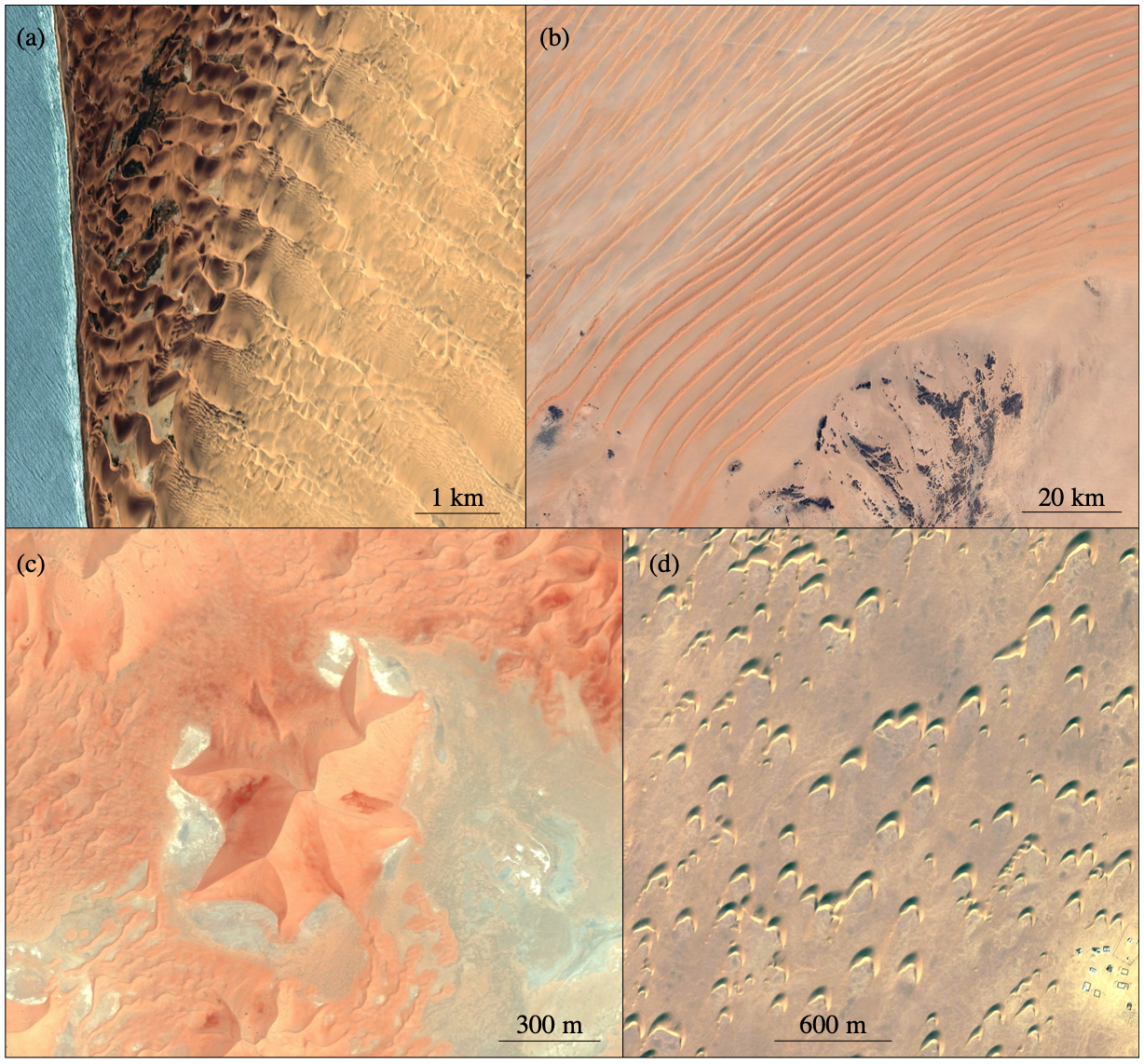}
\end{center}
\caption{Various dune types on Earth. (a) Barchanoid ridges (left, along the coast) and transverse long-crested dunes (right, inland) in Angola (16.336$^{\rm o}$ S, 11.836$^{\rm o}$ E, November 2002, credit: Image Landsat / Copernicus). Dunes inland are transverse to the resultant direction \cite{Cour24}. (b) Linear dunes in the Rub al-Khali desert (16.817$^{\rm o}$ N, 45.869{$^{\rm o}$} E, December 2010, credit: Image Landsat / Copernicus). (c) Star dune in the Rub al-Khali desert (20.388$^{\rm o}$ N, 54.331{$^{\rm o}$} E, April 2007, credit: Maxar Technologie). (d) Field of barchan dunes in Occidental Sahara (27.20$^{\rm o}$ N, 13.21$^{\rm o}$ W, November 2022, credit: Maxar Technologies).
\label{FigDunes}
}
\end{figure}

To study the fundamental processes at work in dune dynamics, field observations take advantage of remote sensing to explore vast areas, but they are limited to a narrow range of time scales and often lack of sufficiently idealised configurations to specify the initial condition or separate coupled effects. Here we discuss the contribution of experiments in the understanding of dune pattern development, and this paper is constructed as follows. A first theoretical section briefly reviews the growing mode in height associated with a bed instability, and an elongation process in multidirectional wind regimes (Sec.~\ref{theory}). Section 3 describes laboratory experiments, where, in order to cut down space and time scales of aeolian transport processes, dune dynamics is analogously studied in water tanks and channels, where the grains are moved by a water flow. Section 4 is devoted to landscape-scale experiments, i.e., dealing in the field with actual space and time scales of the dune under controlled initial and boundary conditions.

Importantly, sedimentary bedforms also naturally develop in subaqueous environments, such as rivers and sea floors. Although we do exploit here the (partial) analogy between aeolian and subaqueous sediment transport, we will not review the vast literature on sedimentary subaqueous pattern, which includes not only ripples and dunes, but also other types like anti-dunes, chevrons and bars associated with specific geometries and regimes of free-surface channelised water flows \cite{Best05, Carl00, Pars05, Iked83, Cole94, Cole96, Cole03, Baas94, Vend05, Lang07, Ouri09, Chan71, Lisl91, Lanz00, Deva10, Rodr14}. Furthermore, we will not address in detail the sensitivity of sediment transport to environmental parameters (e.g., fluid and grain density, grain size distribution). Instead, we will abstract sediment transport into a few key quantities that control the main features of dune development in all these different environments. Finally, we will not consider cohesion between grains, nor the effect of vegetation. These aspects are mentioned in the conclusion as perspectives.

\section{Dune theory}
\label{theory}
In order to present the general framework of dune pattern formation, it is first necessary to introduce the dimensionless numbers that concern sediment transport by a fluid flow over a granular bed \cite{Dura11a, Vala15, Paht20}. We then review the instability mechanism by which dunes can emerge from a flat bed under a unidirectional flow \cite{Andr02, Kroy02a, Kroy02b, Clau06, Nart09, Char13, Cour15}. We finally discuss the case of a multi-directional flow regime, especially where dunes can elongate over a consolidated substratum \cite{Cour14, Gao15a, Luca15, Rozi19}. These theoretical considerations serve as a basis for the development and interpretation of the experiments, both in the laboratory and in the field.

\subsection{Key quantities for sediment transport}
A fluid flow over a granular surface can set the grains into motion when the shear stress it exerts on the bed exceeds a certain threshold. Three main dimensionless parameters controlling the corresponding sediment transport can be built \cite{Dura12}. Considering grains of bulk density $\rho_p$ and diameter $d$ in a fluid of density $\rho_f$ and kinematic viscosity $\nu$, we define the density ratio $s$, the Shields number $\Theta$ to rescale the bed shear stress $\rho_f u_*^2$, and the Galileo number $\mathcal{G}$ to rescale the viscosity. They write:
\begin{equation}
s = \rho_p/\rho_f, \qquad \Theta = u_*^2 / u_g^2 \, , \qquad \mathcal{G} = u_g d / \nu \, ,
\label{sThetamathcalG}
\end{equation}
where $g$ is the gravity acceleration, and $u_g = \sqrt{(s -1) gd}$ a reference velocity scale. As shown by numerical simulations \cite{Dura12}, $s$ essentially governs the transition from bedload ($s$ of the order of a few units, the moving grains keep contact with the bed) to saltation ($s$ larger than a few tens, the grains experience jumps and make dissipative rebounds when colliding with the bed). $\Theta$ quantifies the strength of the flow, especially with respect to the threshold value $\Theta_t$ below which transport vanishes. $\mathcal{G}$ distinguishes viscous ($\mathcal{G} <1$) and turbulent ($\mathcal{G} > 1$) transport conditions, which affects particle drag and flow properties in general. Another important -- yet not independent -- quantity is the Reynolds number based on the grain size $\mathcal{R}_d = d u_*/\nu$. It tells whether the flow over this granular bed is in the smooth or rough regime \cite{Clau17, Jia23}.

Sediment transport can then be summarised into three key quantities. The first one is the already mentioned threshold $\Theta_t$. The corresponding threshold velocity is then $u_t \propto u_g$. The second quantity is the saturated flux $q_{\rm sat}$ at given flow strength, i.e. the vertically integrated maximal amount of sediment that such a flow is capable to transport per unit time and width in the reference homogeneous and steady configuration. Defined as a volumic flux, its scaling factor is then $d u_g$. The actual flux $q$ needs space to relax towards this saturated value if it changes (because wind speed increases for example). The length scale associated to this relaxation is the third quantity and is called the saturation length $L_{\rm sat}$ \cite{Saue01, Andr02, Kroy02a, Andr10, Lamm17, Selm18}. How $u_t/u_g$, $q_{\rm sat}/(d u_g)$ and $L_{\rm sat}/d$ vary with $s$, $\Theta$ and $\mathcal{G}$ is a current topic of research, especially in the context of sediment transport in planetary conditions \cite{Kok12, Berz17, Andr21, Paht23}. However, the review of such studies is out of the scope of the present review. In fact, for the subject of this paper, generic behaviours are enough knowledge. For example, $q_{\rm sat}$ is an increasing function of $\Theta$, typically linear in the case of aeolian (terrestrial) saltation: $q_{\rm sat} \propto d u_g (\Theta - \Theta_t)$ \cite{Unga87, Rasm96, Andr04, Kok12, Crey09, Paht21}. Also, the saturation length is significantly smaller for subaqueous bedload than for aeolian saltation, typically in proportion of $s$ \cite{Andr10}.

\subsection{Sand bed instability mechanism}
\label{sec: bi}
Here we briefly derive the linear stability analysis of a granular erodible bed submitted to a constant fluid flow, which contains the fundamental processes that contribute to the emergence and the growth of dunes \cite{Andr02, Clau06, Char13}. We denote as $h(x,t)$ the bed elevation profile, which depends on the direction of the flow $x$ and time $t$. Bed elevation is related to the sediment flux $q$ by mass conservation, which writes
\begin{equation}
\partial_t h + \vec{\nabla} \cdot \vec{q} = 0.
\label{MassBalance}
\end{equation}
In the simple case of unidirectional flow considered here, the second term in Equation~\ref{MassBalance} reduces to $\partial_x q$. A flux increase ($\partial_x q>0$) is associated with erosion ($\partial_t h<0$), and inversely for flux decrease and sand deposition. As discussed in the previous subsection, $q$ relaxes towards its saturated value $q_{\rm sat}$, and only the latter is an explicit function of the shear stress applied by the flow on the bed. A simple choice for this relaxation is to take it as a first order process, which we write \cite{Andr02}:
\begin{equation}
L_{\rm sat} \partial_x q = q_{\rm sat} - q,
\label{FluxRelaxation}
\end{equation}
but other options are possible \cite{Saue01, Lamm17, Selm18}.

\begin{figure}[htb]
\begin{center}
\includegraphics[width=0.44\linewidth]{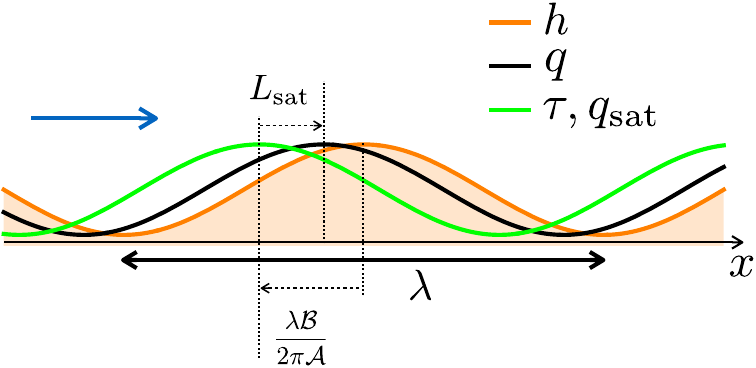}
\hfill
\includegraphics[width=0.54\linewidth]{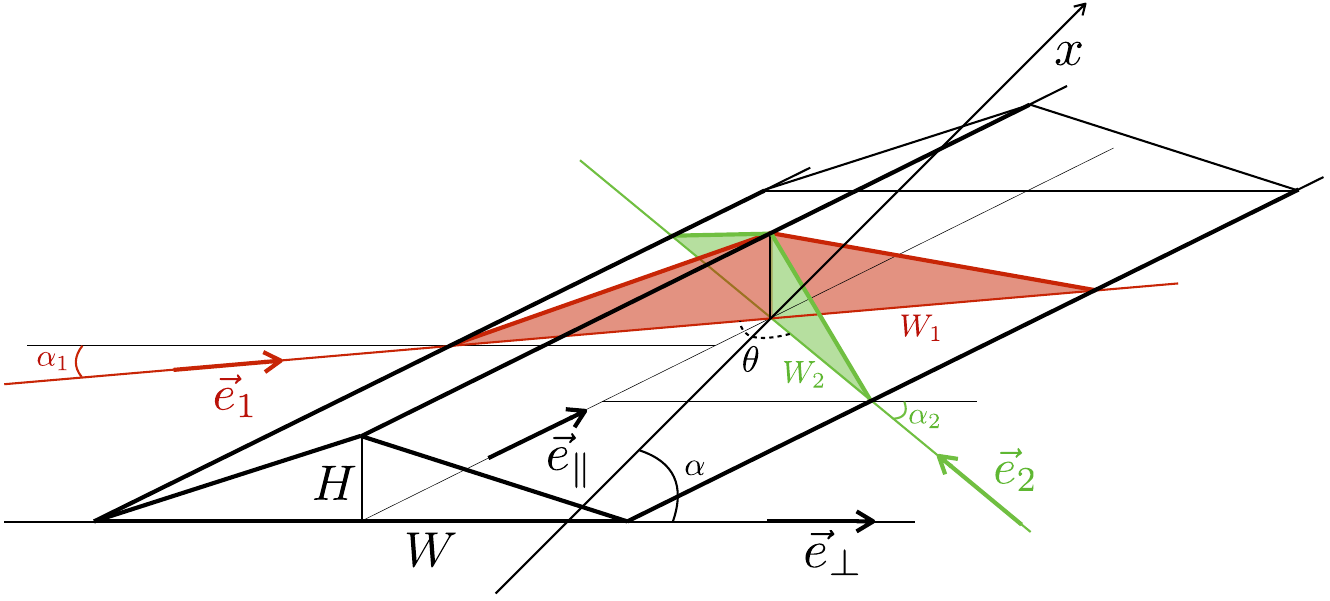}
\end{center}
\caption{
Left panel: Schematic illustration of the upstream shift $\sim \lambda \mathcal{B}/(2 \pi \mathcal{A})$ of the basal shear stress $\tau$ with respect to the bed elevation profile $h$, and the spatial lag $L_{\rm sat}$ between the saturated flux $q_{\rm sat}$ and the actual flux $q$. Flow is from left to right, in the $x$-direction.
Right panel: Schematics of two winds blowing alternately over a symmetric triangular dune of height $H$ and width $W$. The normal- and parallel-to-crest directions are aligned with $\vec{e}_\perp$ and $\vec{e}_\parallel$, respectively. The primary wind 1 (resp. the secondary wind 2) blows along the direction $\vec{e}_1$ ($\vec{e}_2$), making an angle $\alpha_1$ ($\alpha_2$) with respect to $\vec{e}_\perp$. It sees an apparent dune width $W_1$ ($W_2$). The reference $x$-axis is along the bisector of the two winds, which make a divergence angle $\theta$. The orientation $\alpha$ of the dune is defined as the angle made by $\vec{e}_\perp$ with respect to $x$.
\label{FigTheory}
}
\end{figure}

Another piece of this analysis is the hydrodynamic response of the flow with respect to a perturbation of the bed elevation, in comparison to the reference case of a flat bed. This part writes simplest with a decomposition of the bed elevation profiles in Fourier modes. As we are interested in this subsection in the linear regime of the dune instability, we can consider the generic form for the bed perturbation (here in complex notation, real part is understood):
\begin{equation}
h(x,t) = \zeta e^{i(kx-\omega t) + \sigma t} \, ,
\label{hsinus}
\end{equation}
where $k$ is the wavenumber of the perturbation (or equivalently the wavelength $\lambda = 2\pi/k$), and $\zeta$ its amplitude. We will assume $k\zeta \ll 1$. The goal is to derive the dispersion relation that relates the growth rate $\sigma$ and the phase velocity $c=\omega/k$ as a function of $k$. The bed shear stress in response to such a perturbation writes, at linear order in $k\zeta$:
\begin{equation}
\tau = \rho_f u_*^2 \left[ 1 + kh (\mathcal{A} + i \mathcal{B}) \right],
\label{tausinus}
\end{equation}
where $\mathcal{A}$ and $\mathcal{B}$ are the in-phase and in-quadrature dimensionless shear stress coefficients. Experiments \cite{Zilk77, Fred88} and theoretical \cite{Benj59, Hunt88, Weng91, Char00, Kroy02b, Four10} analyses show that these coefficients are typically both positive, and of the order of unity. This means that the basal shear stress is in phase advance with respect to the bed elevation (Fig.~\ref{FigTheory}). In other words, the flow is maximum upstream of the ridge, by a distance $\sim \lambda \mathcal{B}/(2 \pi \mathcal{A})$. Importantly, $\mathcal{A}$ and $\mathcal{B}$ are functions of $k\nu/u_*$, and also differ in the rough ($\mathcal{R}_d > 10$) and smooth ($\mathcal{R}_d < 1$) regimes in a range of wavenumbers around $10^3 \, u_*/\nu$, associated with an anomalous response \cite{Char13, Clau17}. Here, for simplicity, we assume that $\mathcal{A}$ and $\mathcal{B}$ are constants, which essentially corresponds to the case of aeolian dunes in the rough regime.

The perturbation in shear stress (Eq.~\ref{tausinus}) can be transposed to the saturated flux for any given transport law relating $q_{\rm sat}$ to $\Theta$, and a similar linear expression is obtained -- strictly speaking, this is true when the transport threshold can be neglected, and things are slightly more complicated otherwise \cite{Char13, Gada19}. Importantly, $\tau$ and $q_{\rm sat}$ are in phase, but, as a consequence of the spatial relaxation (Eq.~\ref{FluxRelaxation}), $q$ lags downstream of $q_{\rm sat}$ by $L_{\rm sat}$ (Fig.~\ref{FigTheory}). Assuming that $L_{\rm sat}$ is a constant \cite{Andr10}, it is possible to solve these expressions in order to determine the growth rate of a bed perturbation of wavenumber $k$,
\begin{equation}
\sigma = \frac{Q}{L_{\rm sat}^2} \, (k L_{\rm sat})^2 \, \frac{\mathcal{B} - \mathcal{A} k L_{\rm sat}}{1 + (k L_{\rm sat})^2} \, ,
\label{sigmaofk}
\end{equation}
where the reference flux $Q = \tau \partial_\tau q_{\rm sat}$ is here defined as the susceptibility of the saturated flux with respect to the shear stress \cite{Andr02, Elbe05, Clau06, Char13}. The numerator of this function, which determines the sign of $\sigma$, reflects the competition between a hydrodynamic destabilising process associated with the upstream shift of the shear stress (positive contribution to $\sigma$, larger $\mathcal{B}/\mathcal{A}$) and a stabilising process associated with the downstream lag of the flux (negative contribution to $\sigma$, larger $L_{\rm sat}$). In particular, this function shows a cut-off length, or a minimal wavelength, below which all perturbations are stable: $\lambda_{\rm min} = 2\pi L_{\rm sat} \mathcal{A}/\mathcal{B}$. The most unstable wavelength is for $\lambda_{\rm max} \simeq 2 \lambda_{\rm min}$. A metre-scale value of the saturation length for aeolian saltation \cite{Andr10} gives $\lambda_{\rm max}$ on the order of $15$--$30$~m, in good agreement with observations \cite{Elbe05, Gada20a, Delo20, Lu21}. The typical growth time of the instability is $L_{\rm sat}^2/Q$, i.e. typically of the order of a few hours during a storm event \cite{Elbe05}, or on the order of a few weeks if we consider more regular intermittent wind conditions in terrestrial deserts \cite{Lu21}.

Another important point is that this bed instability is convective, and one can also derive the velocity $c$ of the bed perturbation of wavenumber $k$ from the linear analysis. It reads:
\begin{equation}
c = \frac{Q}{L_{\rm sat}} \, k L_{\rm sat} \, \frac{\mathcal{A} + \mathcal{B} k L_{\rm sat}}{1 + (k L_{\rm sat})^2} \, .
\label{cofk}
\end{equation}
For small values of $k L_{\rm sat}$, that is for large enough bed wavelength, $c \simeq Q \mathcal{A} k$, a scaling proportional to $Q$ and inversely to bedform size that is recovered for developed dunes and sometimes named Bagnold's law for propagative dunes (barchan in particular) \cite{Bagn41,Elbe08}. As seen below, the fact that larger dunes migrate slower is an important process for the transverse instability (Sec.~\ref{sec: elong}).

Given the assumptions under which it has been derived, Equation~\ref{sigmaofk} provides a theoretical dispersion relation that can be generalised to all environment according to few generic quantities, namely $Q$, $L_{\rm sat}$, $\mathcal{A}$ and $\mathcal{B}$. The transition from the meter scale of saturation length in a context of aeolian saltation to a millimetre scale (predicted, but yet unmeasured) for underwater bedload allows for a downscaling of pattern emergence from tens of meters and weeks in terrestrial deserts to centimetres and a tens of seconds in an underwater environment, while preserving the primary physical processes. These numbers are of course for the linear regime of the instability. Once the pattern grows in height, non-linearities soon appear. They are first associated with the hydrodynamic response relating $\tau$ to $h$ \cite{Zilk77, Buck84}, then with the development of an avalanche slip face on the downstream side of the bumps \cite{Lu21}, and later to the coarsening dynamics characterised by the rearrangement/elimination of pattern defects \cite{Day18} and interactions between bedforms behaving as individual objects \cite{Elbe05, Hers05, Vala11, Gao15b, Jarv22}. However, the working hypothesis for the laboratory experiments is that the downscaling methodology remains valid for non-linear processes, including the long-term dynamics of dune fields.

\subsection{Multi-directional flow regime}
\label{sec: multidirectional}
The above linear analysis can be generalised in the case of a multi-directional flow regime, which we discuss here for simplicity in the bidirectional case \cite{Gada19}. As schematised in Figure~\ref{FigTheory}, the primary and secondary flows are associated with time duration $t_{1,2}$, reference fluxes $Q_{1,2}$ and directions along unitary vectors $\vec{e}_{1,2}$. We call $\theta$ the divergence angle between these two directions, take the $x$-axis as their bisector, and define the transport ratio $N=Q_1 t_1/(Q_2 t_2)>1$ between the two flows \cite{Cour14, Rubi87, Rubi90, Reff10}. The period of flow reorientation $t_1 + t_2$ is assumed to be small in comparison to the characteristic growth time of the dune. The existence of this time period associated with the flow change provides an additional length scale to the problem. We can define $W_R = \sqrt{Q_1 t_1}$. It sets the typical streamwise size below which a dune loses the memory of its past shape after the primary wind has blown, and thus does not integrate the entire wind regime. In what follows, we consider that $W>W_R$.

\begin{figure}[htb]
\begin{center}
\includegraphics[width=0.44\linewidth]{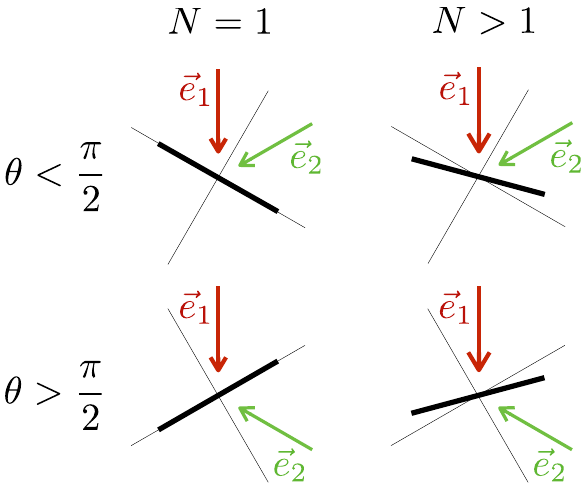}
\end{center}
\caption{
Schematics of the dune orientation when the wind divergence angle $\theta$ is smaller (upper row, dune mostly perpendicular to the resultant) or larger (lower row, dune mostly along the resultant) than $\pi/2$. The selected orientation tends to be more perpendicular to the dominant wind $1$ as the transport ratio $N$ increases (right column).
\label{FigOrientation}
}
\end{figure}

In this case, the most unstable mode not only selects a preferred wavelength $\lambda_{\rm max}$ but also an orientation angle $\alpha_{\rm max}$ of the dune pattern with respect to the $x$-direction. These two quantities depend of course on $N$ and $\theta$ \cite{Gada19}. 
If $N$ is close to one, $\alpha_{\rm max}$ shows a transition from an orientation roughly transverse to the resultant flux $\vec{q}_{\rm sat}$ when $\theta < \pi/2$ (i.e., $\alpha_{\rm max}$ close to $0$) to an orientation roughly aligned with the resultant flux when $\theta > \pi/2$ (i.e., $\alpha_{\rm max}$ close to $\pi/2$), see Fig.~\ref{FigOrientation}. Both winds contribute to the dune elevation regardless of which dune side they blow on, which explains the change in dune orientation with $\theta$. By contrast, $\lambda_{\rm max}$ is less dependent on $N$ and $\theta$.

These results for dune orientation derived in the limit of emerging bedforms with a small aspect ratio ($k\zeta \ll 1$) \cite{Gada19} are consistent with those obtain by a geometric approach \cite{Cour14}, which entirely reduces the role of hydrodynamics on the estimation of the flow speed-up at the crest \cite{Jack75}, and neglects sediment transport relaxation (the spatial lag $L_{\rm sat}$). For transverse dunes of width $W$ and finite height $H$ lying on a flat bed, the flux at the crest writes $q^c = Q (1 + \gamma)$, where the flux-up factor $\gamma$ is proportional to the dune aspect ratio $H/W$ (a typical value $\gamma \simeq 1.7$ for an aspect ratio around $1/12$). In the linear analysis for a unidirectional wind, we can derive $\gamma \propto k\zeta \sqrt{\mathcal{A}^2 + \mathcal{B}^2}$, where $k\zeta$ is indeed proportional to the aspect ratio of the transverse bed perturbation. In the considered bimodal flow regime, introducing the angle $\alpha_{1,2}$ between the flow and the normal-to-crest direction, the apparent dune width exposed to the flow is $W_{1,2} = W/\cos\alpha_{1,2}$, so that the flux at the crest is 
\begin{equation}
\vec{q}^c_{1,2} = Q_{1,2} \left(1 + \gamma \cos\alpha_{1,2} \right) \vec{e}_{1,2}.
\label{qcrest12}
\end{equation}
Their contributions add up in
\begin{equation}
\vec{q}^c = \frac{t_1}{t_1+t_2} \vec{q}^c_1 + \frac{t_2}{t_1+t_2} \vec{q}^c_2 = \vec{q}^c_\perp + \vec{q}^c_\parallel,
\label{qcresttot}
\end{equation}
where $\vec{q}^c_\perp$ and $\vec{q}^c_\parallel$ are the normal- and parallel-to-crest components of the mean transport rate, respectively. With the approximation that the growth rate associated with each flow is $\sigma_{1,2} \simeq q^c_{1,2}/(H W_{1,2})$, the most unstable mode corresponds to the dune orientation for which the sum $\sigma_1 + \sigma_2$, i.e. $q^c_\perp$, is maximum -- also referred to as the gross bedform-normal rule \cite{Rubi87, Cour14}. The resulting selected angle is in good agreement with the one obtained from the linear theory \cite{Gada19}, emphasising the key process of dune growth in height by sand transport in the direction perpendicular to the crest.

\subsection{Elongation mechanism and migration on sediment-starved beds}
\label{sec: elong}
Dunes may also form on consolidated beds that are partially devoid of mobile sediment. This configuration is far from those for which the aforementioned linear stability analysis can be applied. Instead of a periodic dune pattern emerging as the result of erosion and deposition of a surface fully covered with sand, dunes develop as isolated objects on the non-erodible bed. As the variability of wind regimes increases, from a unimodal to a multimodal distribution of wind orientation, these dunes may take on a crescentic shape, such as barchans or asymmetric barchans, or a long shape, including linear and seif dunes \cite{Part09, Reff10, Tani12, Cour14, Gao15a, Luca15, Tsoa16} as well as the individual arms of star dunes \cite{Zhan12}. The elongation of these arms or linear dunes is driven by the sediment transported along the crest ($q^c_\parallel >0$), when the winds alternatively hit both sides of the crest. As seen from the above equations, this process, controlled by aspect ratios, is scale free. When these dunes do not migrate, their orientation is that of the main flux and the net flux perpendicular to the crest vanishes: $q^c_\perp = 0$, as well predicted by the geometric approach of \cite{Cour14}. As for the bed instability mechanism under multidirectional flow conditions (Sec.~\ref{sec: multidirectional}), elongation only occurs if the size of the dune is greater than $W_R$, in order to be able to integrate all flow contributions. 

The elongation dynamics of a dune from a fixed sand source has been studied by numerical simulations, where the wind regime is bimodal and the sediment injection rate $J_{\rm in}$ is constant \cite{Rozi19}. The dune grows from the source area without migrating. If $J_{\rm in}$ is large enough, the dune eventually reaches a steady state, with a roughly triangular cross-section of constant aspect ratio as width and height decrease linearly from the source area. The length of the dune is proportional to the ratio $J_{\rm in}/q_{\rm out}$, where $q_{\rm out}$ is the outflux that escapes from the dune uniformly at all cross-sections. At the tip of the dune in the direction of elongation, the section is successively swept by flows impacting from either side of the crest. The authors of \cite{Rozi19} suggest that the tip size is controlled by the its cut-off length $\lambda_{\rm min}$, i.e. related to $L_{\rm sat}$ again. Another possibility is that the dune ends where the transverse instability (see below) can break it, i.e. where its width is on the order of $W_R$.
Elongating dunes are thus strongly linked to their sediment supply, and they often develop from large dunes or at the edge of sand seas. In these cases, the resulting dune field displays a typical spatial periodicity. Numerical simulations have shown that the corresponding wavelength can be inherited from the bed instability developing on the sediment layer, which serves as governing boundary or initial conditions \cite{Cour14, Gada20}.

A last generic configuration is that of dunes that are not attached to a source but can migrate over such a consolidated bed. The simplest case is that of a sand bar of height $H$ transverse to a unidirectional flow. Their migration velocity follows Bagnold's scaling in $Q/H$, associated with mass conservation at the downwind avalanche slipface \cite{Bagn41, Dura11a}. The migration of the dunes induces an instability that tends to break the transverse structure into smaller dunes whose length and width are both of the order of $H$ \cite{Part11,Guig13}. This instability, called transverse instability, results from the positive feedback loop between the migration rate of the longitudinal sections of the dune and the transverse sediment fluxes: lower sections tend to migrate faster than higher ones, curving the avalanche slipface, which redirects the sand towards the highest parts, thus amplifying the phenomenon. This is also what is likely to happen when dunes under multidirectional wind regimes are smaller than $W_R$, which prevents them to elongate.

\section{Laboratory-scale experiments}
\label{labscale}
The minimum size of aeolian dunes, on the order of 10 m in length on Earth, make them difficult to reproduce in laboratory-scale experiments. The alternative is to study subaqueous sedimentary bedforms, which can be downscaled to a length of about a centimetre. Increasing the fluid density (decreasing the density ratio $s$) lowers the threshold fluid velocity for grain entrainment and increases the fluid inertia relative to grain inertia, which consequently reduces the saturation length and the corresponding minimum size for bedforms under water compared to the aeolian case.

The details of sediment transport in subaqueous conditions, typically bedload when the Shields number is not too large, are of course different than those of aeolian saltation. But although the key parameters such as the saturation length $L_{\rm sat}$ and the reference sediment flux $Q$ scale differently with respect to grain and fluid characteristics \cite{Dura12}, the physical background of the dune theory (Sec.~\ref{theory}) is discussed without referring to any particular transport law. A caveat, however is that, as opposed to air flows over saltating grains for which the regime is typically rough, water flows over granular surfaces are more likely to be in the smooth regime, where, depending on grain size and bedform wavelength, a specific hydrodynamic response can occur, which affects the coefficients $\mathcal{A}$, $\mathcal{B}$ \cite{Char13, Clau17}. In this case, the viscous length $\nu/u_*$ comes as an additional relevant length scale for the minimum size for bedforms. Nevertheless, like aeolian dunes, these subaqueous sedimentary bedforms form from the positive feedback loop between the bed topography, the fluid flow, and the sediment transport, with analogous generic processes at play for their formation and dynamics.

\begin{figure}[t]
\begin{center}
\includegraphics[width=0.95\linewidth]{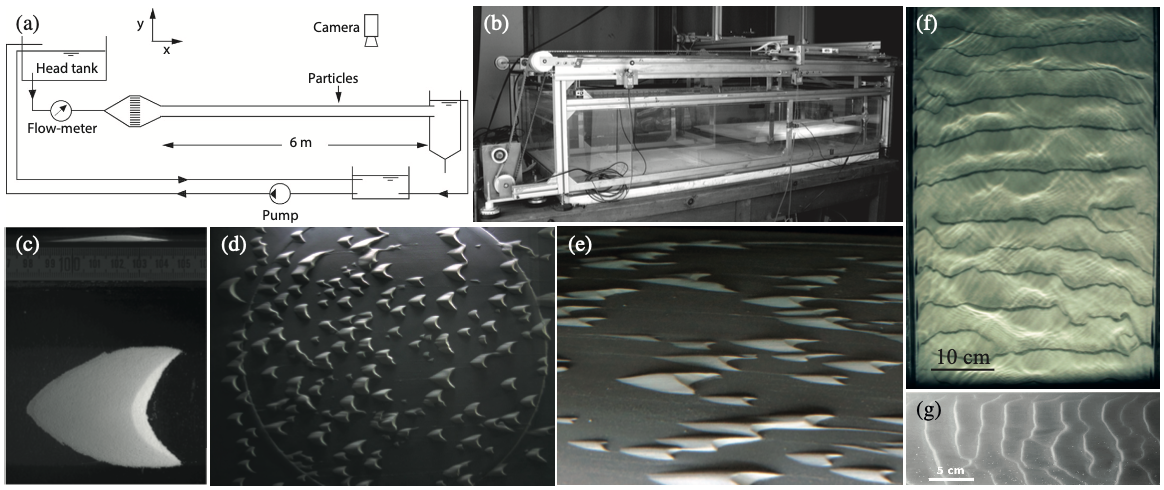} 
\end{center}
\caption{
Different experimental setups, barchan (crescentic) and periodic transverse dunes in unidirectional flow. (a) Sketch of the closed water channel used in \cite{Fran11}. The flow is driven by gravity from the head tank. A convergent and honeycomb system eliminates large turbulent eddies. A tank at the end of the channel collects sand grains and a pump refills the head tank. The channel section is 12 cm wide and 6 cm high. (b) Photograph of the setup used in \cite{Reff10}. A tray (80 cm wide and 90 cm long, here on the right of the tank) oscillates in a water tank (2 m long, 1 m wide, and filled with 50 cm of water). The translation motion is asymmetric. (c) Barchan dune in setup (a) \cite{Fran11}. Flow is from left to right. A barchan dune has a characteristic crescent shape with two horns (or arms) pointing in the dowstream direction. It has a gentle slope ($\sim \, 0.1$) on the upstream side and an avalanche face on the lee side. (d) \& (e) Field of barchan dunes in setup (b) (Photo credit: S. Courrech du Pont). The disk is 70 cm wide. Flow is from left to right. (f) Periodic transverse dunes in an open (free surface) channel setup. Flow is upward. Photo credit: A. Valance and L. Guignier. (g)  Periodic transverse dunes in a closed channel, from \cite{Lang07}. Flow is from right to left. In (f) and (g), dunes are fully developed and exhibit avalanche faces in the lee side. Early pattern is straighter.
\label{FigSetups}
}
\end{figure}

As illustrated in this section, underwater experiments have proven useful in understanding sandy desert landscapes, especially regarding the shape of dunes in response to flow forcing and sand supply conditions, as well as interactions between dunes or with a fixed obstacle. The morphology and migration velocity of subaqueous barchans in experiments have been compared to aeolian barchan on Earth by several authors \cite{Hers02, Endo05, Fran09} and recently reviewed in \cite{Zhan24}. For example, the aspect ratio of height to width is similar and, importantly, both aeolian and subaqueous barchans migrate at a velocity inversely proportional to their height $c \propto Q/H$. However, subaqueous barchans have a slightly larger ratio of width to length and a smaller ratio of horn length to length than aeolian barchans (see Figures \ref{FigDunes} and \ref{FigSetups} for photographs of barchans).

Noteworthy, in the context of fluvial or oceanographic geomorphology, periodic underwater bedforms of small wavelength (typically smaller than a few tenth of centimetres) are commonly referred to as ripples, whereas larger bedforms (larger than a few meters) are referred to as dunes \cite{Vend13}. Here we consider underwater bedform to be analogous to aeolian dunes with respect to the morphodynamical processes and we will indistinctly use the term `dune' for both of them. As these subaqueous dunes are smaller, the characteristic times associated with their evolution are also smaller, allowing `long time ranges' to be observed in laboratory experiments. This is of particular interest because dunes can integrate flows for long periods of time.

\subsection{Experimental setups}
Subaqueous sand patterns have been studied in flumes, which are either pipes \cite{Al-l08,Ouri09}, open \cite{Mant78,Rubi90,Endo04}, or closed channels \cite{Lang07,Groh08, Fran11,Baci20,Assi23,Yang24}, or using an oscillating tray \cite{Hers05, Cour14} (Fig.~\ref{FigSetups}). The confined pipe geometry is not ideal for studying dune morphodynamics as observed in natural environments, but grain or powder entrainment and pattern formation in such a configuration \cite{Ouri09} are relevant to industry \cite{Al-l08}.

Water channel with a rectangular section is the most commonly used device. The flow can be driven by pumps, propellers, or by gravity, which does not produce noise as engines do. Divergent-convergent devices combined to honeycomb help to minimise turbulent eddies. In open channels, the flow can generate unwanted waves at the free surface. Closed channels seem more appropriate and facilitate the use of gravity to drive the flow. The flat top surface also simplifies visualisation.  With transparent walls, water channels enable PIV to characterise the flow with or without dunes \cite{Lang07, Char12} and to follow the paths of individual grains \cite{Alva17}. Some channels are intentionally narrow to study two-dimensional dunes \cite{Lois05, Groh08, Baci20}. These setups are appropriate to study the development of periodic dunes from the destabilisation of a sand bed or single isolated dunes in unidirectional flows. Regarding interactions between dunes, care must be taken when extrapolating results to three-dimensional dunes. Although some experiments have been conducted in large flumes \cite{Rubi90}, the width of the channel is typically between 10 and 20 cm in laboratory \cite{Lang07, Fran11, Endo04}, which limits the number of dunes in the cross-section or the study of dunes in multidirectional flows, as is often the case in Earth deserts where wind directions change with seasons \cite{Cour15}. To study dunes in multidirectional flows, the sand bed is placed on a horizontal turntable that rotates to change the direction of the sand bed with respect to the water flow \cite{Reff10, Tani12}. The width of the setup then limits the length of the dune field.

A way to increase the dimension of the field and to avoid the need for a large flow rate is to use an oscillating tray, which can also be a turntable \cite{Hers02, Reff10}. The oscillating tray's back-and-forth motion is asymmetric, with acceleration and velocity above the transport threshold in one direction and below the threshold in the other. At the end of the fast motion, the tray slowly decelerates to avoid backflow in the tray's reference frame. The repetition of this oscillating motion simulates a unidirectional flow over the bed. The main advantage of this setup is the size of the field that can be studied, a disk of 70 cm in diameter in \cite{Reff10}. Since the tray is at rest twice during the motion cycle, this setup also facilitates controlled sand supply, which is an important parameter when studying dune fields. The main drawbacks are that these experiments take longer than in flumes, and that the intermittent flow and sand transport are difficult to characterise.

In all setups, grains are glass, zirconium, or ceramic beads of diameter ranging from 50~$\mu$m to 1~mm and flow velocities typically range from 20 to 70~cm/s. These values give dimensionless numbers on the order of $s \simeq 2$--$7$, $\Theta \simeq 0.5$--$3$ and $\mathcal{G} \simeq 1$--$100$.

\begin{figure}[t]
\begin{center}
\includegraphics[width=0.9\linewidth]{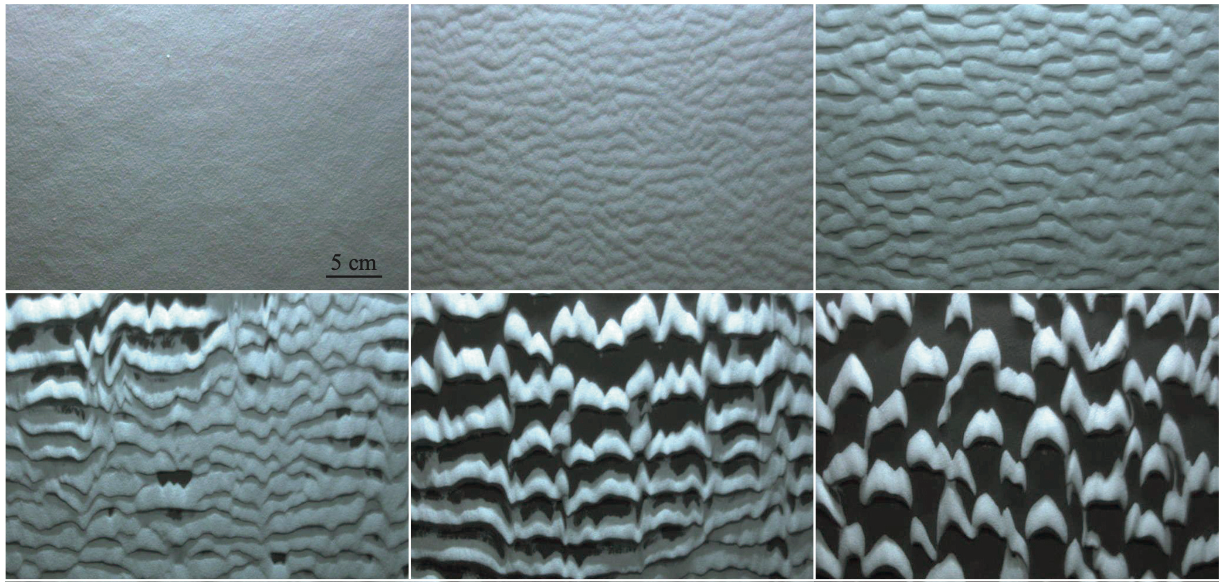}
\end{center}
\caption{Sand bed of limited thickness subjected to a unidirectional flow. Flow is from top to bottom (translated tray setup). Images were taken after none, 40, 180, 410, 560, and 930 tray strokes. Experiment images are from \cite{Reff10}.
\label{FigExpInst}
}
\end{figure}

\subsection{Dune instability}
Experiments have shown that a fully sand covered bed, initially flat, destabilises into a periodic migrating pattern with a well-defined wavelength and velocity. In steady flow, the initial wavelength is about $15 \, \rm{mm}$ for glass beads of $100$~$\mu$m and friction velocity in between 15 and $20 \, {\rm{mm/s}}$ \cite{Lang07}. With such small patterns, the three-step feed-back mechanism (topography / fluid friction velocity / sediment flux) described in Section \ref{sec: bi} has not been directly verified in laboratory experiments -- the corresponding thickness of the boundary layers and saturation length are too small. No linear regime is observed in the development of the instability (exponential growth of amplitude with a constant wavelength) in laboratory setups, probably because it is too short, but rather a coarsening of the pattern from the beginning \cite{Lang07, Ouri09, Reff10, Cour14}. A linear regime has however been observed during $150$~s in a natural river with a bedform wavelength of about $9$~cm for grain of diameter $0.3$~mm and a shear velocity of $4$~cm/s \cite{Four10}. 

\subsection{Dunes in multidirectional flows and different sand supply conditions - the two dune growth mechanisms}
The relationship between wind regimes and dune type, shape and orientation is at the heart of dune studies, not least because dunes and their deposits are proxies for past winds. Most experimental studies focus on understanding equilibrium shapes and how dunes respond to changes in wind regime or sand supply conditions, the two main factors controlling free dune type (free dunes are dunes for which wind and sand transport are not constrained by other factors such as obstacles or vegetation \cite{Pye90, Cook93, Livi96,Cour24}). Here we discuss a selection of experiments, which together with numerical simulations ({\it{e.g.}}, \cite{Part09, Nart09, Dura10, Zhan12}), have shown that dune shape is controlled by its dynamics, which can be decomposed in the three processes of growth in height, migration and elongation \cite{Cour24}. The relative balance of these three dynamics is mainly controlled by flow regime and sediment supply and availability \cite{Cour24}.

\begin{figure}[t]
\begin{center}
\includegraphics[width=0.5\linewidth]{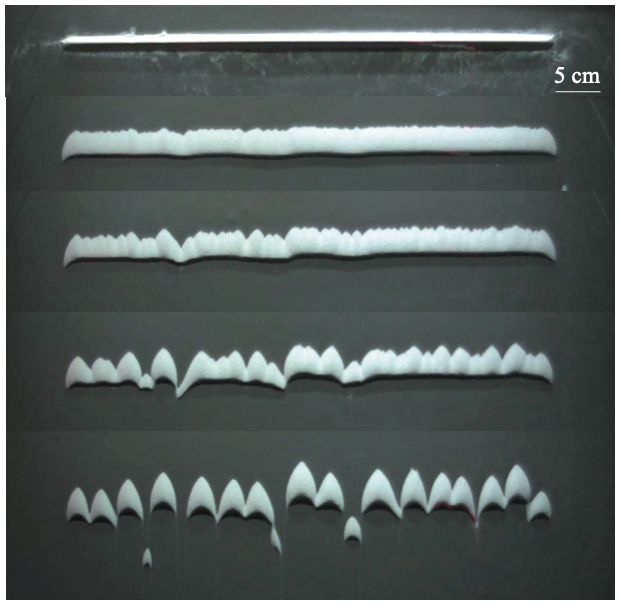}
\end{center}
\caption{Instability of an isolated transverse ridge migrating on a starved bed. Flow is from top to bottom (translated tray setup). Images were taken after none, 40, 100, 200, 400, and 800 tray strokes. Images are from \cite{Reff10}.
\label{FigExpTrans}
}
\end{figure}

\begin{figure}[p]
\begin{center}
\includegraphics[width=0.9\linewidth]{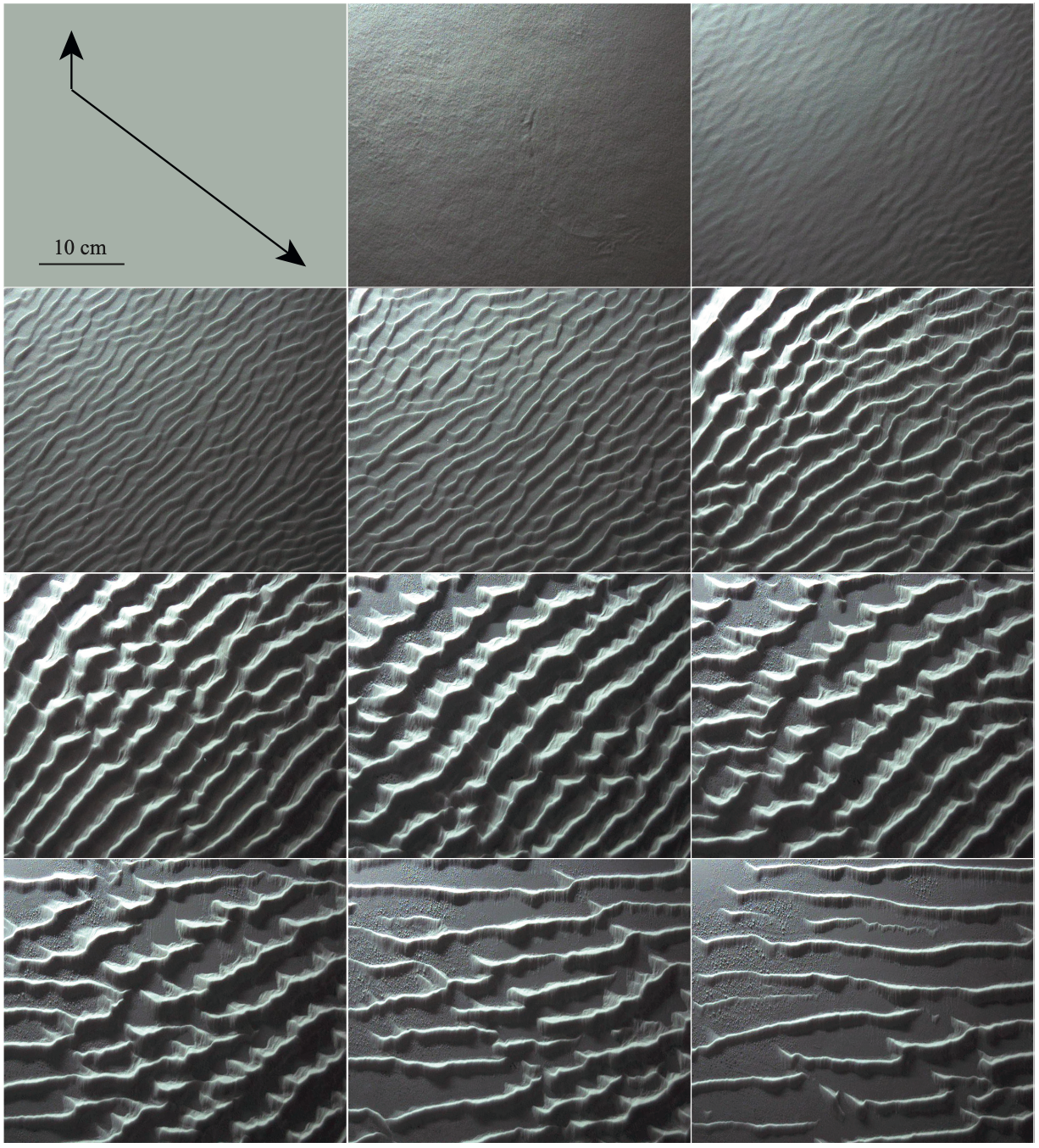}
\end{center}
\caption{Sand bed of limited thickness subjected to a periodic bidirectional flow (translated tray setup). First panel shows the flow regime. The two flow directions form an angle $\theta = 130^\circ$. The tray alternatively translates 10 times in the primary flow direction and 2 times in the secondary flow direction; the transport ratio $N$ between the primary and the secondary flows equals 5. Images were taken at the end of the upward secondary flow after none, 1, 3, 7, 12, 16, 30, 40, 50, 65, and 85 cycles (one cycle counts $10+2=12$ tray strokes). Photo credit: S. Courrech du Pont, from experimental set-up of \cite{Cour14}.
\label{FigExpTwomodes}
}
\end{figure}

\begin{figure}[p]
\begin{center}
\includegraphics[width=0.9\linewidth]{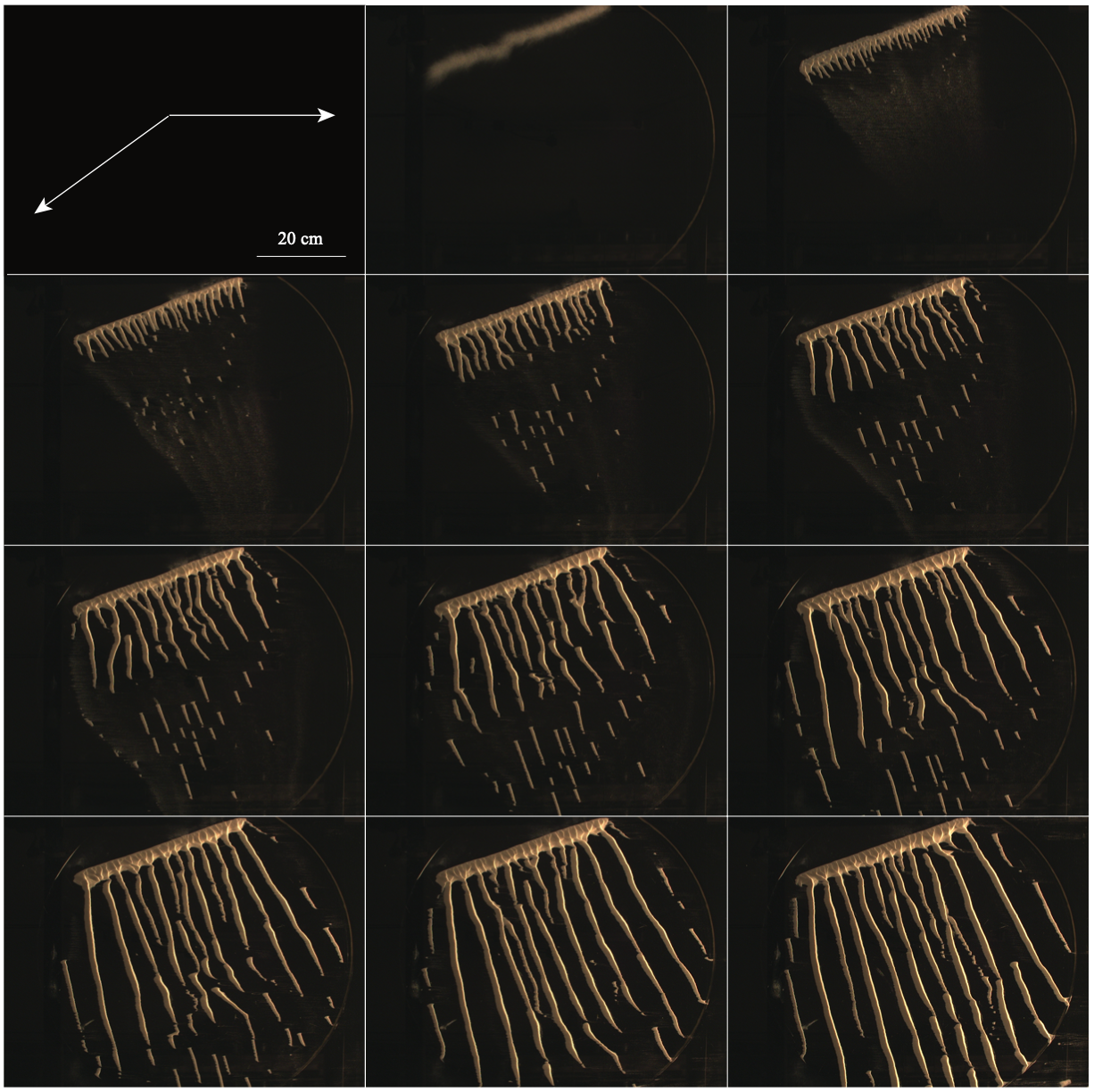}
\end{center}
\caption{Dune growth and elongation from a line source of sand in a periodic bidirectional flow (translated tray setup). Sand is periodically supplied along a line (at top left). First panel shows the flow regime: the two flow directions form an angle $\theta=144^\circ$ and the transport ratio between the two flows is $N=1$ (symmetric). Images were taken at the end of the rightward flow after none, 10, 20, 40, 60, 80, 120, 160, 200, 260, and 400 flow cycles (one cycle counts $3+3=6$ tray strokes). Photo credit: S. Courrech du Pont.
\label{FigExpElong}
}
\end{figure}

Under unidirectional flow and on a fully covered sand bed, dunes grow continuously as periodic long ridges perpendicular to the flow direction (transverse dunes), from the excavation of the underlying substrate and coarsening. When the thickness of the sand bed is limited, the bed gets progressively starved and the transverse dunes break into rows of barchans (Fig.~\ref{FigExpInst}, \cite{Reff10}). This change in dune morphology (long transverse dunes to barchans) corresponds to a change in dominant dynamics (growth in height to migration). On a starved bed, a perturbation in dune height leads to a perturbation in migration velocity (lower parts migrate faster than higher ones). The subsequent sinuosity of the dune crest leads to lateral sediment fluxes that amplify the perturbation (Fig.~\ref{FigExpTrans}, \cite{Reff10, Part11, Guig13}). Figure~\ref{FigDunes}a shows a spatial transition between dunes on completely covered and partially starved sand beds in the field.  

In multidirectional flow regimes, dunes may integrate the different flow sequences (Sec.~\ref{sec: multidirectional}) and each of them may contribute to the development of the dune. Dunes on a fully covered bed in bidirectional flow regimes have been studied experimentally in a large flume by Rubin and Ikeda \cite{Rubi90} (first experiments dealt with the orientation of impact ripples in bidirectional winds \cite{Rubi87}) and later by Reffet {\it{et al.}} \cite{Reff10} and others \cite{Cour14} using the translated tray setup (Fig.~\ref{FigExpTwomodes}). These experiments have shown that dune orientation maximises the gross bedform-normal transport, {\it{i.e.}}, the rate of growth rate in height (Sec.~\ref{sec: multidirectional}). Subsequently, dunes can be transverse, oblique or longitudinal with respect to the mean sand flow direction. This major experimental result has allowed to quantitatively use dune orientation to constrain prediction of wind regimes \cite{Anth96, Fent14, Ewin10}.

In bidirectional flow regimes, when the bed becomes starved, the periodic dunes either break into barchans (which may be asymmetric \cite{Tani12}) or break and reconnect to form periodic dunes with a different orientation (Fig.~\ref{FigExpTwomodes}, \cite{Cour14}). These experiments show that long dunes and periodic patterns can have two different orientation depending on the bed condition (fully covered or starved), and that a mechanism stabilises long dunes on starved beds in multidirectional flow, which is dune elongation. This second growth mechanism (the first being growth in height) can generate long linear dunes on starved beds, a common pattern on Earth (Fig.~\ref{FigDunes},b \cite{Cour14}) and probably on Titan \cite{Char15}) from sand sources (Fig.~\ref{FigExpElong}, \cite{Cour14}.

The space of parameters controlling the dune type has only been partially explored experimentally (in unidirectional or bidirectional flow regimes only), star dunes (Fig.~\ref{FigDunes}d), for example, have not yet been reproduced in experiments. More complex flow regimes and varied boundary conditions need to be addressed in order to obtain the aimed phase diagram of dune patterns.

\begin{figure}[p]
\begin{center}
\includegraphics[width=0.9\linewidth]{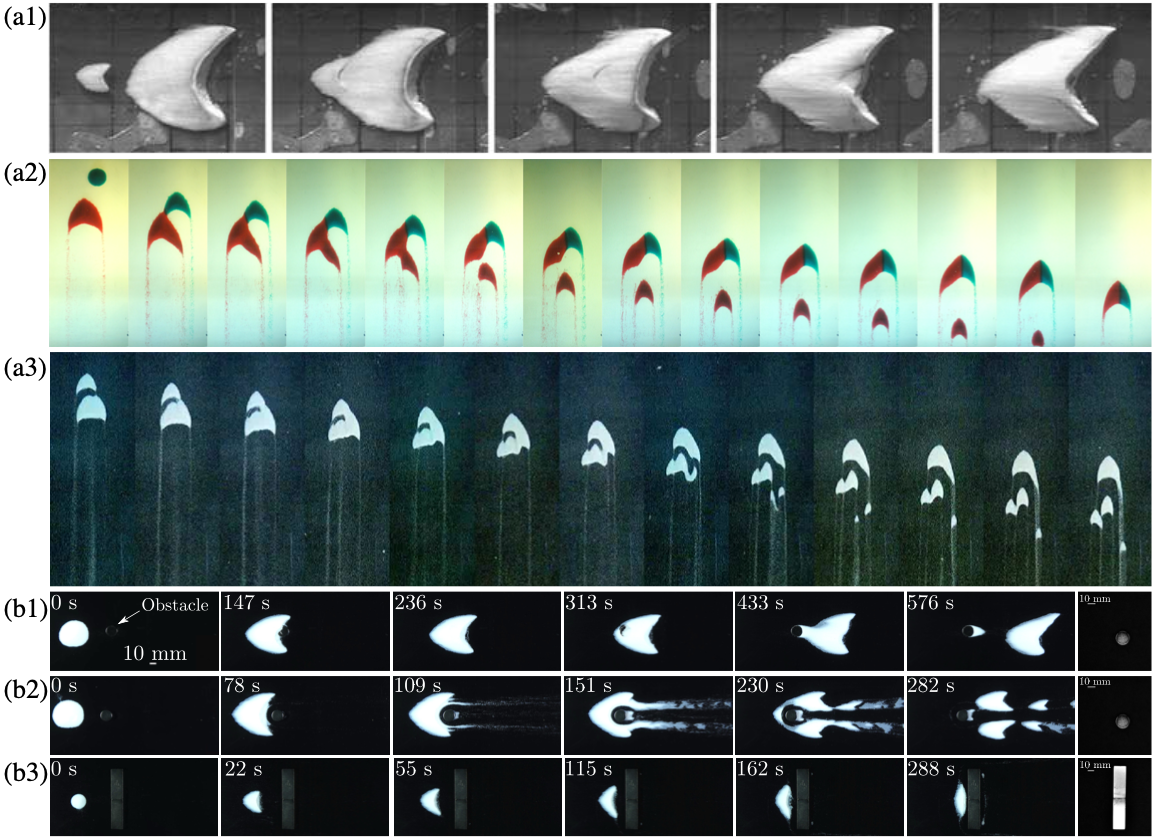}
\end{center}
\caption{Barchan dune interactions. (a1-3) Time-lapses of barchan collisions. Depending on the size ratio and alignment between the two barchans, the two barchans merge (a1),  sand is redistributed into two barchans (a2), or into a larger number of dunes (a3). (a1) Flume setup, flow is from left to right and view in images is 12 cm high, from \cite{Endo04}. (a2) Translated tray setup. Flow is from top to bottom, view in images is 19 cm wide, from \cite{Hers05}. (a3) Translated tray setup, flow is from top to bottom, view in images is 8 cm wide, credit: S. Courrech du Pont \cite{Cour15}. (b1-2) Barchan dunes interacting with an obstacle. The dune can either pass over the obstacle (b1), bypass the obstacle (b2), or get stuck (b3). The flow is from left to right (flume setup) and the obstacle is either a cylinder 18 mm in diameter and 2 mm high (b1), a cylinder 18 mm in diameter and 8 mm high (b2), or a rectangular block 1 cm high, 10 cm wide, and 2 cm long (in the flow direction). Note the sand trapped in the lee of the obstacle in (b1). (b1-3) figures are from \cite{Assi23}. 
\label{FigInteract}
}
\end{figure}

\subsection{Dune interaction and autogenic processes}
Dunes highly interact with their environment. Dune interactions can be divided into three categories: dune interaction with obstacles, dune interaction with each other, and dune interaction with vegetation. Experiments are particularly important for dune interactions because numerical models of dunes only partially reproduce the complexity of the flow and, in particular, the wake that develops downwind of the dune.

Barchan interactions with obstacles has recently been investigated by Assis {\it{et al.}} (Fig.~\ref{FigInteract}, \cite{Assi23}). In their study obstacles have sizes comparable to the size of dunes. They have shown that depending on the different aspect ratios between the dune and the obstacle, the different aspect ratios of the obstacle, and the flow velocity, the dune will either pass over the obstacle, bypass the obstacle, or get stuck and form an echo dune. During the process, sand may be trapped in the lee side of the obstacle and form a lee (or shadow) dune. These studies are particularly important for risk prevention and management of human infrastructure in desert environments.
  
Under a unidirectional wind, dunes on starved beds are migrating barchan dunes. Barchan dunes loose sand from their horns and can capture free sand flux. Numerical models have shown that the balance between gain and loss corresponds to an unstable equilibrium. The output flux depends little on the size of the dune and is nearly constant, whereas the gain is proportional to the width of the barchan if the free flux is homogeneously distributed \cite{Hers04}. An isolated barchan should either shrink or grow. In deserts, barchan are not isolated, but are numerous and interact (Fig.~\ref{FigDunes}d). Their interactions appear to regulate their size and their arrangement within the field \cite{Bagn41, Elbe05}. Two types of interactions have been identified corresponding to long-range and short-range sand exchange. (i) Sand lost by a barchan from its horn is captured by a barchan downstream, driving dune arrangement in echelon \cite{Bagn41, Elbe08}. (ii) Smaller barchans migrate faster than larger ones, leading to dune collisions. Experiments have shown that dune collisions can result in dune merging, sand redistribution between the two dunes, or formation of new dunes, depending on the relative size and alignment between impacting and impacted dunes (Fig.~\ref{FigInteract}, \cite{Endo04, Hers05, Assi20}). Dunes merge if the impacting dune is much smaller than the impacted one. If the impacting dune is larger, the flow perturbation downstream the impacting dune locally increases the fluid velocity experienced by the impacted dune enough to cut it into pieces. The impacting dune may merge with one or more of the newly formed dunes \cite{Endo04, Hers05}. Dune splitting during collisions has been shown to regulate dune size using mean-field approaches \cite{Hers05, Dura09}. Following the experimental identification of collision types and parameters, a more complete phase diagram of binary collisions was developed in numerical simulations \cite{Dura09, Jarv23}. These collision rules were then used to simulate the dynamics of large groups of barchans with agent models, demonstrating that barchan collisions can not only regulate the size of barchans, but also result in the self-organisation of the dune field in corridors \cite{Hers04, Geno13, Worm13, Geno13b, Geno16}.

The other important type of interaction is the interaction with vegetation. The coupling between sand transport, dunes, and vegetation is also dynamical. Vegetation can grow on dunes ,and dunes can kill the vegetation, which remobilises the underlying sand \cite{Hesp02, Dura06, Hesp13}. This coupling is still a challenge for laboratory experiments, but is of primary importance because it is relevant for most coastal deserts and for vegetated dunes that cover large areas in China and Australia.

\section{Landscape-scale experiments}
\label{lanscapescale}
The concept of landscape-scale experiments is a novel approach to geomorphology that differs from the usual field instrumentation or remote sensing analysis, and is particularly suited to modern issues in aeolian research and dune physics. Indeed, although the minimum size of the dunes and their migration rate remain beyond the operational capacity in wind tunnels, the temporal and spatial scales associated with dune growth mechanisms appear to be accessible in the field, enabling the validation and quantification of the processes that contribute to the formation of dune patterns in their natural environment. Given the extreme conditions typically encountered in arid deserts, along with the specific properties of wind regimes, these in-situ experiments necessitate the integration of logistical considerations for the acquisition of long-term measurements. These challenges have been met for $6$~years in the Tengger Desert thanks to a Franco-Chinese collaboration combining theoretical approaches with field experience.

\begin{figure}[p]
\begin{center}
\includegraphics[width=0.8\linewidth]{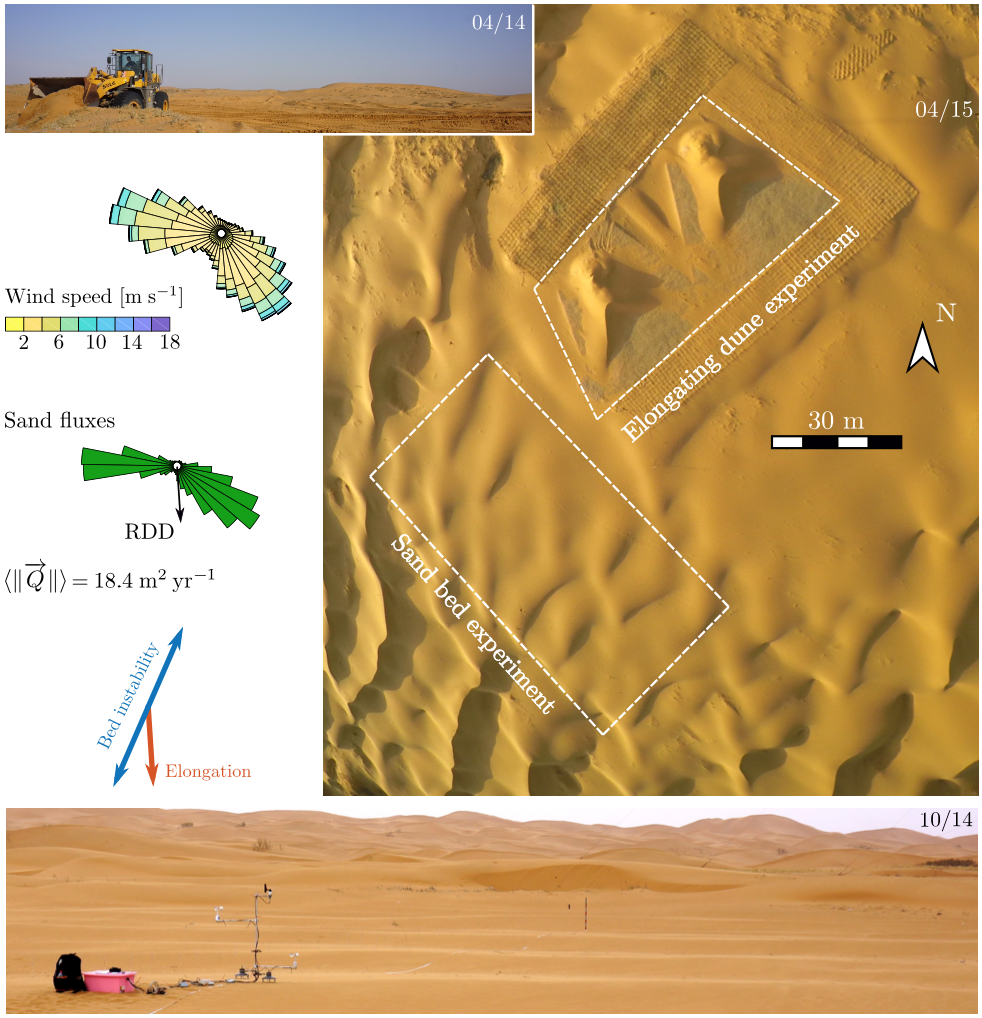}
\end{center}
\caption{
Landscape scale experiments in the Tengger Desert, China (37.560$^{\rm o}$N, 105.033$^{\rm o}$E). Two independent experiments are dedicated to the elongation and bed instability mechanisms. Wind and transport roses point in the direction of the flow. Arrows show dune orientations predicted from wind data. The bottom picture shows wind measurements at different heights along a wind-parallel transect during the development of the dune instability.
}
\label{fig: landscale_1}
\end{figure}

\subsection{Experimental set-up in the field}
\label{sec: setup}
Working on surfaces large enough for elementary dunes to be observed, the main objective of the landscape-scale experiments is to quantitatively characterise the development of bedforms in a natural aeolian environment using controlled initial and boundary conditions as well as a continuous monitoring of the surface winds and climatic conditions that contribute to aeolian transport. Such control, inherent in laboratory experiments, is the main difference with field observations, and also helps to limit the impact of some of the heterogeneities (e.g., vegetation, grain-size distribution) often encountered in desert areas.

The experimental site is located in the southernmost border of the Gobi Desert, which has a typical continental monsoon climate with less than $180~{\rm mm}$ of annual precipitation. The local wind regime is clearly bimodal (Fig.~\ref{fig: landscale_1}), with a dominant wind in the southeast direction flowing mainly in spring when the Siberian-Mongolian high pressure cell weakens, and a secondary wind in the west direction flowing mainly in summer due to the East Asian monsoon. Figure~\ref{fig: landscale_1} shows that this bimodal wind regime gives rise to a bimodal transport regime, which may be modeled by a divergence angle of $\theta \simeq 149^{\rm o}$ and a transport ratio of $N \simeq 2$.

Field experiments have been continuously conducted over $4$~years from October~2013 to November~2017. To focus on the bed instability and elongation mechanisms described in Sections~\ref{sec: bi} and \ref{sec: elong}, two main experiments have been performed (Fig.~\ref{fig: landscale_1}). The elongating dune experiment started in October 2013 by placing two conical sand heaps $2.5$ and $3\,{\rm m}$ high on a flat gravel bed isolated from the incoming sand by straw checkerboards. To allow for dunes to increase their size and compensate for their loss, sand has been added $10$~times, from May 2014 to September 2016, always in the same places, to rebuild the original sand heaps. To the southwest of the elongating dune experiment, pre-existing dunes were also levelled by bulldozer in April 2014 to form a flat rectangular bed $100\,{\rm m}$ long and $75\,{\rm m}$ wide. The main axis of this sand bed experiment dedicated to the bed instability mechanism was aligned with the direction of the primary wind.

More than 20 ground-based laser scans were performed to regularly measure the evolution of surface elevation in the sand bed and the elongating dune experiments over $4$~years. The point density varied from $470$ to $2\,370$ points per ${\rm m}^2$ with a centimetre height accuracy. Throughout the experiments, wind data from local meteorological towers and an airport located $10\,{\rm km}$ east from the experimental dune field are used to calculate the saturated sand flux on a flat sand bed and at the dune crests according to the formalism described in \cite{Cour14}. In the experimental area, the mean grain size is $d \simeq 190\;\mu{\rm m}$ and the threshold shear velocity for aerodynamic entrainment of sand grains is $u_t \simeq 0.23\;{\rm m\,s}^{-1}$ \cite{Lu21}.

Additionally, a zone of bare sand has been preserved between the two sand heaps of the elongating dune experiment. There, a rectangular sand slab $12\,{\rm m}$ long, $3\,{\rm m}$ wide and $20\,{\rm cm}$ high was built downwind of the gravel bed and the straw checkerboard according to the direction of the prevailing wind. The changes in elevation of this sand berm before and after a wind event was used to estimate the saturation length $L_{\rm sat}$. Finally, the wind response to topography, as encoded in the theory by means of coefficients $\mathcal{A}$ and $\mathcal{B}$ (Eq.~\ref{tausinus}), was measured with anemometers moved along a known (approx. sinusoidal) dune elevation profile (Fig.~\ref{fig: landscale_1}).

\begin{figure}[p]
\begin{center}
\includegraphics[width=0.8\linewidth]{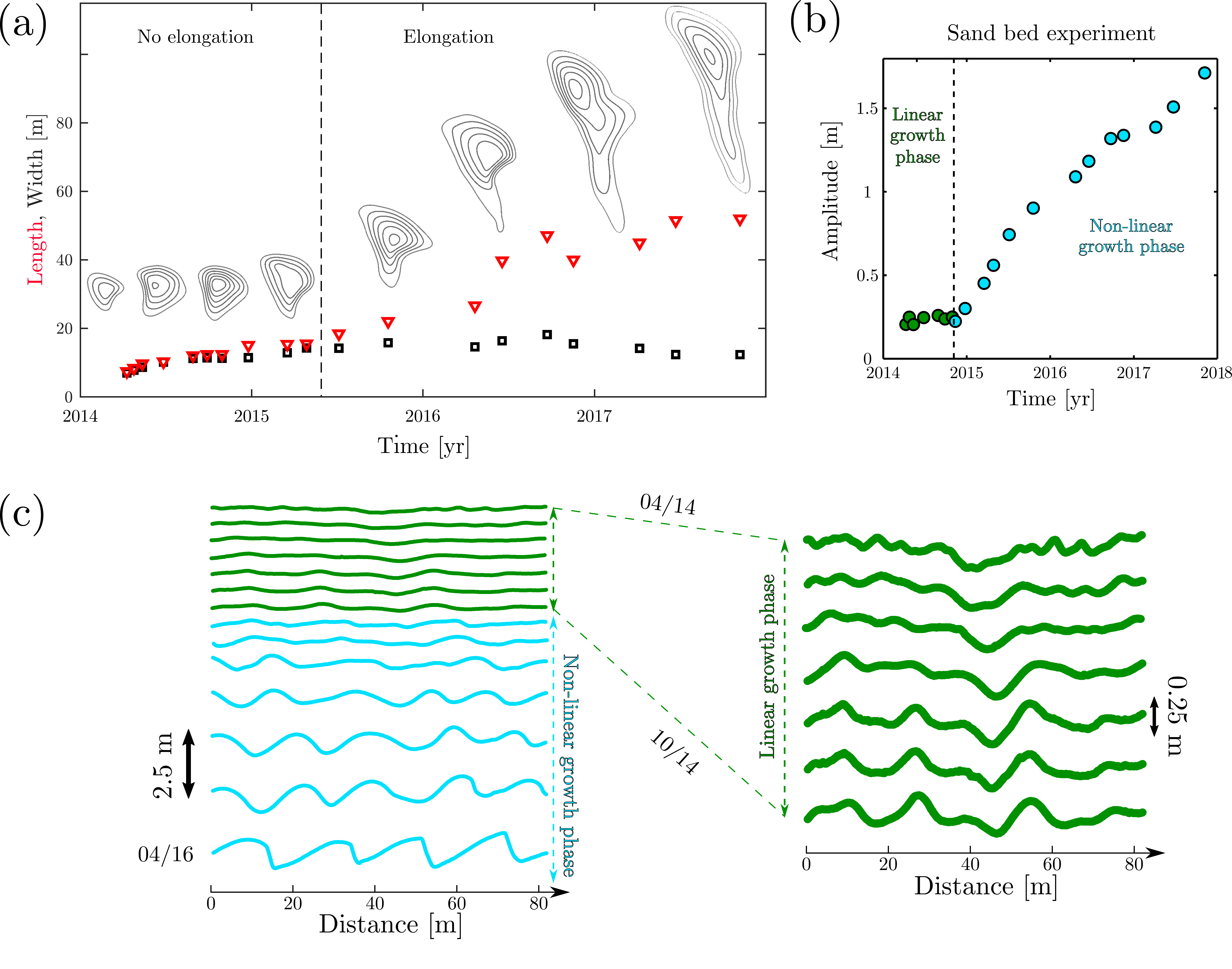}
\end{center}
\caption{
The evolution of bedforms in landscape scale experiments.
(a)
Length and width of the southwest dune in the elongating dune experiment with respect to time. Contour lines show the dune at different times.
(b) 
Average amplitude of the topography variations $\delta h = 2\sqrt{2}(\langle h^2\rangle-\langle h\rangle^2)^{1/2}$ in the sand-bed experiment with respect to time.
(c)
Evolution of topography along a transect of the sand-bed experiment from the initial flattening. The colour distinguishes the linear (green) and non-linear (blue) regimes of dune growth.
}
\label{fig: landscale_2}
\end{figure}

\begin{figure}[p]
\begin{center}
\includegraphics[width=0.8\linewidth]{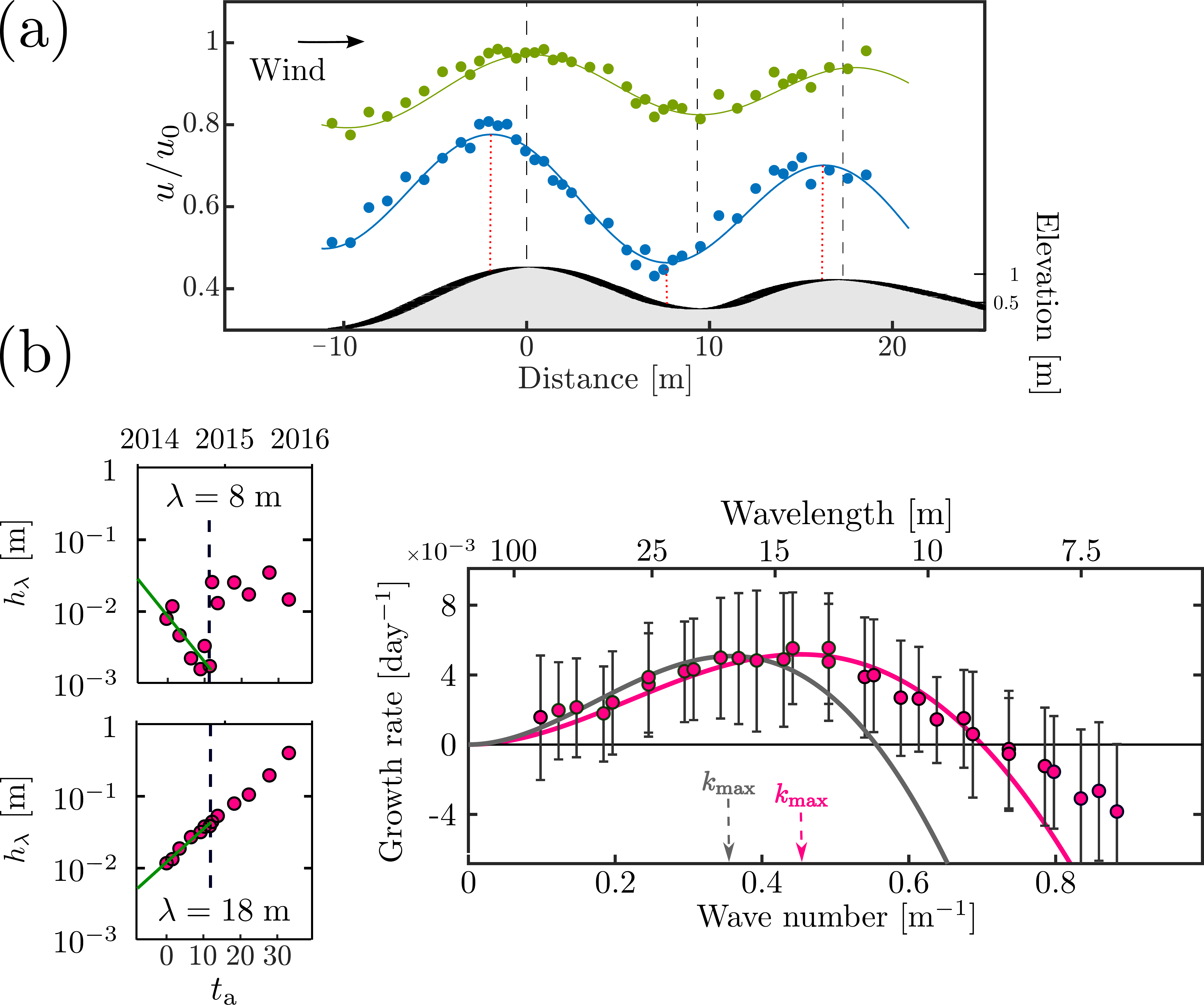}
\end{center}
\caption{Experimental and theoretical dispersion diagrams from field data.
(a)
Elevation and normalised wind speed according to distance along a dune profile (blue and green for anemometers at $4$ and $100\;{\rm cm}$ height, respectively). The wind speeds are normalised by a wind speed measured by a fixed reference anemometer. The average upwind shift gives $\mathcal{A} \simeq 3$ and $\mathcal{B}\simeq 1.5$. 
(b)
Experimental dispersion relation for the linear regime of the dune instability (red points). It is derived from topographic data only, using the exponential variation of the amplitude of different wavelengths from 10 April to 30 October 2014 (see examples for wavelength of $8$ and $18\;{\rm m}$). Solid lines are dispersion relations using Eq.~\ref{sigmaofk} and the best fit to the data $\{L_{\rm sat},\, \mathcal{A},\, \mathcal{B}\}=\{0.7~m,\,1.96,\,0.96\}$ (red), or the same parameters measured independently in the field $\{L_{\rm sat},\, \mathcal{A},\, \mathcal{B}\}=\{0.95~m,\,3,\,1.5\}$ (gray). The most unstable wavelengths $\lambda_{\rm max}=2\pi/k_{\rm max}$ are equal to $14.6$ and $18.5\;{\rm m}$, respectively.
}
\label{fig: landscale_3}
\end{figure}

\subsection{Characteristic length scales in landscape-scale experiments}
\label{sec: lse_scales}
The initial volumes of the two sand heaps of the elongating dune experiment were insufficient to integrate the entire period of wind reorientation, and they rapidly took a dome dune shape under the effect of the first wind reversals \cite{Gao18}. Subsequently, seasonal winds reshaped the dunes, resulting in the formation of crescentic barchans with varying orientations (Fig.~\ref{fig: landscale_2}a). The regular sand supply combined with the exchange of sediments between the two piles led to the growth of the pile to the northeast. This pile developed a southeast-facing slip face, a longer southern arm, and gradual adopted an asymmetric barchan shape. Then, the width of the dune remained constant, while its southern arm continued to elongate at an angle of less than $10^{\rm o}$ to the resulting sand flux in response to the alternating winds (Fig.~\ref{fig: landscale_1}a). This illustrates not only the dependence of dune shape on dune size, but also the existence of a threshold dune size for the elongation mechanism.

In the sand-bed experiment, the working surface of $7,500\;{\rm m}^2$ was initially flattened down to differences between minimum and maximum elevations of less than $25\;{\rm cm}$, with an average residual variation around $\delta h \approx 20\;{\rm cm}$ in April~2014 (Fig.~\ref{fig: landscale_2}c,d). Aeolian transport rapidly smoothed all traces left by the levelling process, and migrating bedforms of different wavelengths start to develop. While a dominant wavelength of approximately $15\;{\rm m}$ started to be visible in August 2014, it was only in October 2014 that the amplitude of these growing bedforms exceeded the initial $\delta h$ and that slip faces oriented to the southeast began to emerge. The growth continued at a steady rate until the completion of the topographic surveys in October~2017, resulting in a clear periodic dune pattern in October~2016 as the amplitude of the dominant mode increases by almost a factor of five.

Taking advantage of this incipient growth, wind measurements were conducted to quantify the flow perturbation over low sinusoidal bedforms and estimate the aerodynamic coefficients $\mathcal{A}$ and $\mathcal{B}$ (Sec.~\ref{sec: bi}). As predicted by the theory (Fig.~\ref{FigTheory}), Figure~\ref{fig: landscale_3}a shows that the wind perturbation reflects the topography of the underlying incipient dunes. The amplitude of the wind speed perturbation decreases with height above the bed. Compared to the elevation profile, there is an upwind shift in wind speed for the bottom anemometers, but not for the top anemometers, in agreement with the existence of inner and outer layers \cite{Hunt88}. In the near-surface region, here typically for heights of less than $15~{\rm cm}$ (in the inner layer), the upwind shift ($\approx 1 \,{\rm m}$) and the amplitude of the wind speed perturbation were used to get $\mathcal{A}=3 \pm 1$ and $\mathcal{B}=1.5 \pm 0.5$. These values are confirm theoretical expectations \cite{Char13} as well as previous but less precise estimates \cite{Clau13}. Independently, the measurement of the saturation length, derived from the erosion of the sand berm during a storm event in April~2014, gave $L_{\rm sat}=0.95\;{\rm m}$. These values allowed us to compute the theoretical dispersion relation of the dune instability (Eq.~\ref{sigmaofk}, as displayed with the blue curve in Fig.~\ref{fig: landscale_3}b).

Using all topographic surveys of the sand-bed experiment, the elevation profile along different transect are used to estimate the evolution of the contribution of individual modes (i.e., wavelengths) to the overall topography. When expressed according to a dimensionless transport time scale ($\Delta t_{\rm a}=Q\Delta t/L_{\rm sat}^2$), all the individual modes exhibit exponential variation over short times (Fig.~\ref{fig: landscale_3}b). Once again, abrupt changes in growth rate are observed in November~2014, a date at which the exponential rate is no longer ubiquitous. The results of the field experiment not only corroborate the dune instability theory with regard to the initial exponential growth of individual modes, but also identify the transition from the linear to the non-linear regime of dune growth, which occurs for remarkably low values of dune aspect ratio ($\approx 0.03$).

\begin{figure}[p]
\begin{center}
\includegraphics[width=0.9\linewidth]{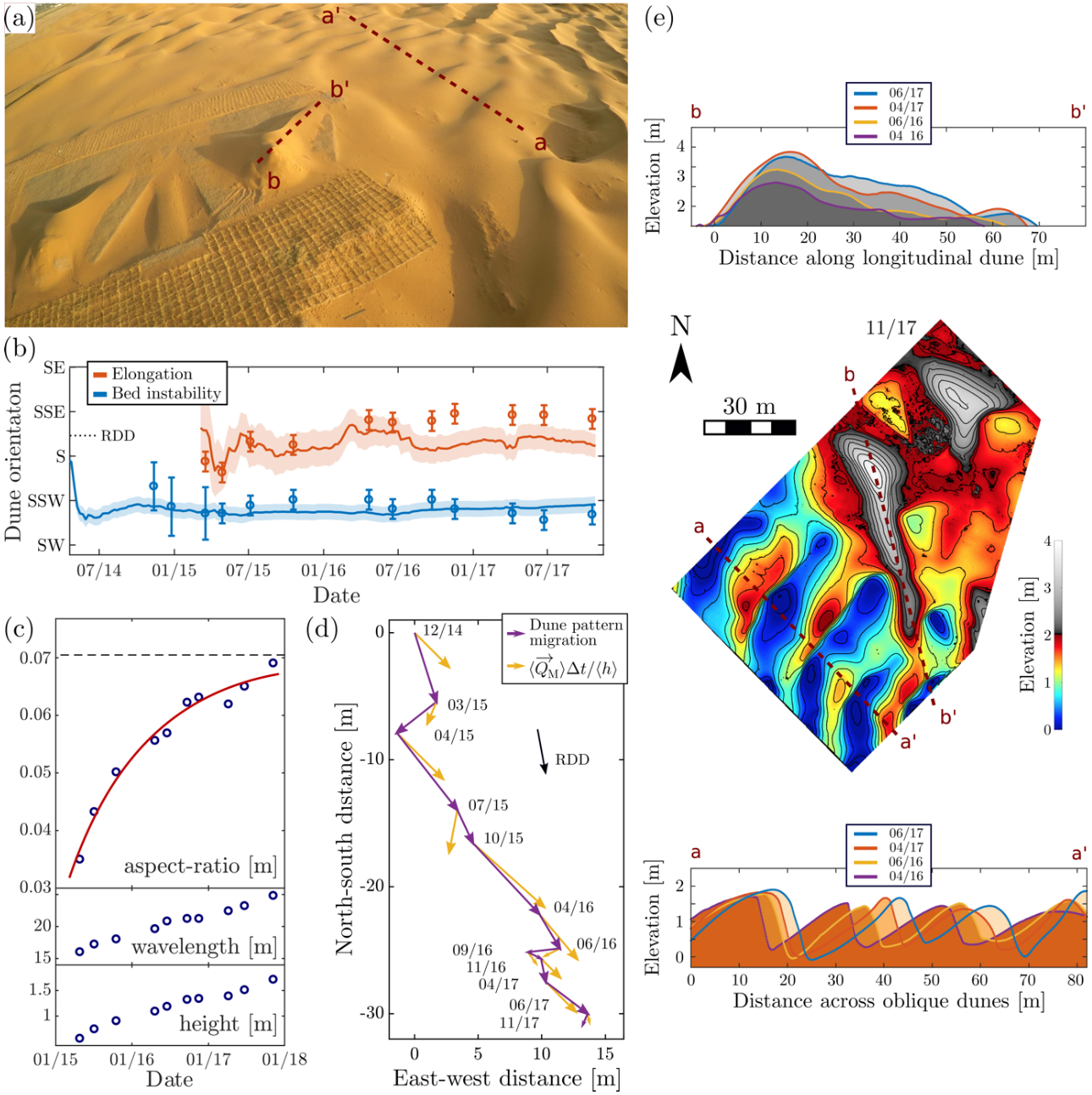}
\end{center}
\caption{Dune dynamics in landscape scale experiments.
(a) 
Direction along which migration and elongation are measured in the field
(b)
Observed (circles) and predicted (lines and shaded area) dune orientations for the dune instability (blue) and the elongation mechanisms (red).
(c)
Evolution of dune aspect-ratio, wavelength and amplitude in the sand-bed experiment.
(d)
Observed (purple) and predicted (yellow) dune migrations in the sand-bed experiments with respect to time.
(e) Elongation (top) and migration (bottom) in the elongating and sand-bed experiments, respectively.}
\label{fig: landdynamics}
\end{figure}

During the linear regime, from April to October~2014, the experimental dispersion relation of the dune instability was obtained by plotting the exponential growth rate of the different modes as a function of their wave number $k$ (Fig.~\ref{fig: landscale_3}b). The growth rate is positive for large wavelengths and negative for small wavelengths, reaching a maximum value for an intermediate wavelength of approximately $15\,{\rm m}$. The presence of clear maximum and the continuous trend from unstable (i.e., growing  waves, $\sigma>0$) to stable regimes (i.e., decaying waves, $\sigma<0$) reflect most of the behaviours predicted by the theory during the linear phase of dune growth. This is demonstrated by the consistency between the experimental dispersion relation, derived exclusively from topographic data, and the theoretical relation, derived from the pertinent physical parameters that were independently measured in the field (Fig.~\ref{fig: landscale_3}b). Other studies confirmed the general agreement between observations and theoretical predictions for dune emergence \cite{Elbe05, Gada20, Delo20}, but, importantly, here it comes from measurements performed on a purposely prepared system, rather than `passive' observations.

\subsection{Dune dynamics in landscape-scale experiments}
\label{sec: lse_dynamics}
In addition to the elementary processes associated with the emergence of dunes, landscape-scale experiments permit the analysis of the long-term dynamics of aeolian bedforms, including dune shape and orientation, as well as the rates at which they grow in height, migrate, and elongate (Fig.~\ref{fig: landdynamics}).

At the time of each topographic survey, the mean transport rate at the crest of dunes can be computed using the wind data acquired in the field from the beginning of the experiments. This allows for a comparison of the observed and predicted (Secs.~\ref{sec: multidirectional},\ref{sec: elong} and \cite{Cour14}) dune orientations, both for the migrating dunes in the sand-bed experiment and elongating dunes. Figure~\ref{fig: landdynamics}b illustrates that these dune orientations are in good agreement with expectations, despite the simplifying assumptions inherent to the model (i.e., triangular dune, no crest reversal). As showed by \cite{Ping14}, the periodic dune orientation in the zone of full sand availability validates the gross bedform-normal rule in the field, as they align with the direction in which the normal component of transport is maximum \cite{Rubi87}. Since they form an angle of $34^{\rm o}$ with the resultant sand flux, they may be described as oblique \cite{Hunt83}. In contrast, the elongating dunes may be considered as longitudinal, since they diverge by only $10^{\rm o}$ from the resultant sand flux \cite{Hunt83}.

During the non-linear phase of dune growth in the sand-bed experiment, Figure~\ref{fig: landdynamics}c shows a constant increase in both dune height and wavelength. The dune aspect-ratio slowly converges toward a steady-state value of 0.07, the typical dune aspect-ratio observed in this area. As the periodic dunes grow in height, the entire pattern also migrates. The migration vectors can be derived from two successive elevation data sets by cross-correlation and compared to the model prediction by computing the sand flux at the crest for the specific time period and dune orientation (Fig.~\ref{fig: landdynamics}d). The observed vectors are consistent with the predicted vectors, both in terms of magnitude and direction, indicating that the model estimate accurately reflects the sand flux at the crest. In the sand bed experiment, dunes eventually reached a height of $2\,{\rm m}$ and migrated at an average speed of $5\;{\rm m}\,{\rm yr}^{-1}$ from June 2016 to June 2017 (Fig.~\ref{fig: landdynamics}e). In the elongating dune experiment, the length of the longest dune increased by more than $10\,{\rm m}$ from April 2016 to June 2017, with a height profile which decreased almost linearly from the sand source area to the tip, in agreement with the predictions of \cite{Rozi19}.

\section{Conclusion and perspectives}
\label{conclu}

The interpretation of dune landscapes, which result from complex interactions (superimposed bedforms, collisions) and integrate wind regimes over long times, can be achieved by identifying the elementary physical processes at work and their associated scales. For this purpose, idealised and controlled experiments, both at laboratory and landscape scales, are an invaluable source of knowledge. A comprehensive range of dunes, encompassing a multitude of shapes and orientations, can be replicated and tracked over time, from their emergence to their interactions and coarsening dynamics. These experiments are also key to assess the models with which these processes are theoretically described, and to quantify their relevant parameters. In turn, the theory provides a framework for understanding the main controlling scales, in particular by analysing the extent to which the processes at work in the morphodynamics of subaqueous laboratory experiments dealing with bedload and centimetre-scale sedimentary bedforms can be analogous to those of large aeolian sand dunes generated by saltation.

This approach of dune dynamics based on processes and scales also gives a quantitative reference and building blocks for numerical models, such as agent-based and cellular automata, to simulate larger dune fields. In particular setting the values of the saturation length $L_{\rm sat}$ and the sand flux $Q$ allows for a quantitative calibration of the results \cite{Nart09}. It also lays the physical foundation for the comparison of dunes in terrestrial and planetary environments \cite{Bour10,Lore14,Luca14,Jia17,Day18,Lapo18,Dura19,Cour24}, for which fluid and sediment characteristics can take values significantly different than those commonly in terrestrial deserts.

Given the central role of the saturation length in dune dynamics, further experiments are required to investigate how $L_{\rm sat}$ depends on environmental parameters, and to relate it to the properties of the fluid and the grains, the flow regime, and the structure of the transport layer. In particular, low pressure conditions (larges values of $s$, the density ratio between the particles and the gas), which are typical of planetary surfaces, must be further explored \cite{Andr21}. In this limit, one expects a smooth hydrodynamic regime to emerge associated with enhanced values of kinematic viscosity as well as a reduced drag force on the grains \cite{Jia17, Dura19, Cour24}. The same is true for the sediment flux $Q$ and the transport threshold $u_t$. This knowledge is essential for using these dune patterns as observational tools to learn more about the flow and sediment properties in places where direct measurements are often difficult to implement. Planetary environments as diverse as those reported on Mars and Titan will thus continue to provide a wide range of dune patterns that will undoubtedly serve to guide the development of new experiments, as well as provide empirical constraints to improve the theoretical parametrisation and description of the processes.

When sufficiently large, dune flanks can destabilise like flat beds \cite{Elbe05}, explaining how a hierarchy of dune patterns develops, from the most unstable mode to the giant dune size, with different generations of dunes interacting with each other \cite{Andr09, Gao15b}. These primary and secondary structures are valuable sources of information on the properties of surface winds over different periods, from seasonal to millennial time scales. In areas where sediment availability varies spatially, the different growth mechanisms can also coexist \cite{Lu17,Lu22}, or the same mechanism can produce dunes with different orientations \cite{Zhan12}. Experiments have not yet addressed superimposed patterns and complex shapes such as star dunes or raked dunes. Further long-term experimental studies are required to investigate dune interaction and coarsening processes in this context of superimposed patterns or for the spatial organisation of dune fields, such as barchan corridors \cite{Elbe08}. Of particular interest is a better understanding of the dynamics by which dune fields eliminate or create defects in the dune pattern, in order to quantify how far from a possible equilibrium state the dune system is \cite{Day18,Marv23}. This long term evolution would also provide a set of constraints on the wind regimes that may have generated these dune patterns \cite{Gada22}, making the inverse problem of characterising past winds from the shape and orientation of dunes potentially feasible \cite{Casc18}.

Finally, dealing with vegetation is an important step for laboratory and landscape scale experiments dedicated to dune dynamics. Coastal zone studies and management over the last few decades have generated a vast amount of data and observations, which have yet to be fully exploited \cite{Cast19}. All this information offers significant potential for the development of new field experiments under controlled conditions in the presence of vegetation \cite{Lapo21}. Meanwhile, growing plants in the laboratory is a difficult time consuming task, and it remains to be determined what the best experimental analogue of vegetation is. A primary objective would be to ascertain some mechanical properties of plants that are relevant to sediment transport and dune morphodynamics. All these experimental efforts should be combined with studies investigating the role of cohesion (e.g., moisture, fine material, sintering, but also coupling to icy surfaces as on Mars), which introduces additional sources of interaction between bedforms, and can modify shear stress and sediment transport, eventually leading to possibly similar feedbacks of those involved by vegetation \cite{Rala22}.

\section*{Acknowlegments}
The authors would like to thank B. Andreotti, M. Baddock, S. Carpy, F. Charru, P. Delorme, S. Douady, O. Dur\'an, H. Elbelrhiti, A. Fourri\`ere, C. Gadal, X. Gao, M. G\'enois, A. Gunn, P. Hersen, J. Iversen, P. Jia, M. Lap\^otre, M. Louge, P. L\"u, A. Lucas, J. Merrison, B. Murray, J. Nield, O. Pouliquen, K. Rasmussen, E. Reffet, S. Rodriguez, O. Rozier, D. Rubin, G. Wiggs, D. Zhang and Y. Zhang for extensive discussions on ripples, dunes and sediment transport.
We are grateful to A. Valance and E. Franklin for sharing images from their experiments.
CN acknowledges the financial support of the French National Research Agency through grants ANR-18-IDEX-0001, and ANR-23-CE56-0008.

\bibliographystyle{elsarticle-num}

\end{document}